\documentclass[a4paper]{article}

\usepackage{INTERSPEECH2020}
\usepackage[table,xcdraw]{xcolor}
\usepackage{multirow}
\usepackage{multicol}
\usepackage[inline]{enumitem}

\title {Users' Perceptions about Teleconferencing Applications \\Collected through Twitter}
\name{Abraham Woubie, Pablo Pérez Zarazaga, Tom Bäckström}
\address{
  Department of Signal Processing and Acoustics, Aalto University, Espoo, Finland }
\email{\{{abraham.zewoudie, pablo.perezzarazaga, tom.backstrom}\}@aalto.fi}

\begin{document} 
 
\maketitle
\begin{abstract}
The COVID-19 outbreak disrupted different organizations, employees and students, who turned to teleconference applications to collaborate and socialize even during the quarantine. Thus, the demand of teleconferencing applications surged with mobile application downloads reaching the highest number ever seen. However, some of the teleconference applications recently suffered from several issues such as security, privacy, media quality, reliability, capacity and technical difficulties.  Thus, in this work, we explore the opinions of different users towards different teleconference applications. Firstly, posts on Twitter, known as tweets,  about remote working and different teleconference applications are extracted using different keywords. Then, the extracted tweets are passed to sentiment classifier to classify the tweets into positive and negative. Afterwards, the most important features of different teleconference applications are extracted and analyzed.  Finally, we highlight the main strengths, drawbacks and challenges of different teleconference applications.

\end{abstract}

\noindent\textbf{Index Terms}: quality, remote work, security, sentiment analysis, teleconference

\section{Introduction}

Due to the worldwide spread of the COVID-19 crisis, large parts of the world were locked-down and many companies, institutions, schools and universities introduced remote working policies. As working from home became common practice, having a reliable video conferencing service also became more important than ever. Having an effective video conferencing service allows  flexibility to work remotely from home. Thus, getting accessible and affordable tools is vital. Consequently, the use of video conferencing applications has surged and these applications (i.e., Zoom, Skype, Microsoft Team, FaceTime, HouseParty, Google Hangouts, Google Duo, Jitsi, etc) were downloaded 62 million times between March $14$ and $21$ \cite{WinNT}.

The above mentioned teleconference services have existed for years, but the changing business scenario and advanced technology features are forcing people into the world of video-chatting. The large number of new users however puts the systems on a test: How well do these platforms cope with the increased number of users and importantly, what kind of problems do users face when starting to use teleconferencing software? This latter question is important from a scientific point of view; we are looking at these applications from a speech processing perspective and if there are substantial deficiencies in such speech processing applications, then it would be the task for the speech processing research community to provide better methods. \emph{By providing better tools for teleconferencing, we, as a speech research community, can help alleviate the negative consequence of the current and future pandemics.} In addition, by developing teleconferencing applications further, many more face-to-face meetings can be replaced by teleconferences even in post-pandemic times. We can thus help in reducing the need for business travel and thereby supporting sustainability goals~\cite{UN}.

We expect that the list of hurdles users could face when using teleconferencing applications is long; they could for example have problems with
\begin{enumerate*}
\item installing software,
\item the network and connection,
\item audio hardware (headphones, microphones etc.),
\item hardware performance (CPU, battery and network capacity),
\item environmental distortions (background noises and reverberation),
\item transmission-related distortions (coding artefacts, transmission delays, packet-loss etc.),
\item navigating the user-interface,
\item social interaction and norms in an unfamiliar environment,
\item privacy and security, etc.
\end{enumerate*}
We however do not have objective information with regard to the prevalence and severity of problems in each area, nor do we have a
comprehensive list of typical problems users face. In news sources, some of the typical problems mentioned include security and privacy,
to the extent that unauthorized participation in teleconferences has been named as \emph{"zoombombing"} following the name of one of the
service providers~\cite{nytimes2020zoombombing}.

Thus, the objective of this study is to find quantitative and qualitative evidence of the performance of teleconferencing applications, and to help focus future 
research efforts in related areas of speech processing. We want to determine which particular types of problems users face and their relative importance. We will focus on all main-stream teleconferencing services applications available, which include:

\begin{itemize}
    \item Zoom is a collaborative cloud-based videoconferencing service that offers various features including online meetings, group messaging services \cite{zoom}. 
    
    \item Microsoft Teams is a chat and collaboration platform for Microsoft Office 365 customers. 
    \item Skype is an application suitable for video conferencing for small teams of up to 50 people including the host. 
    \item FaceTime is a video-calling application designed by Apple for use on the iPhone, iPad, and Mac.
    
    \item Google Hangout supports chatting with up to 150 people, but video calls with only up to 10 participants.
    
    \item Google Duo is a video chat mobile application available on the Android and iOS operating systems. It is also available to use via Google's Chrome web browser.
    
    \item Houseparty is a social networking service that enables group video chatting through mobile and desktop applications. 
    \item Others such as WhatsApp, Signal, FreeConference, Cisco Webex Meetings, GoToMeeting, CyberLink U Meeting,  BlueJeans and  Lifesize.
\end{itemize}

In this work, we are mainly interested in the most widely used video conference applications. However, we did not study Facebook, WhatsApp, Signal, Telegram and other similar applications since their features are much broader and extracting only information related to videoconferencing is challenging.

To collect information about the sentiments of users, we have chosen to source data from the microblogging and social networking site Twitter. Many people post their opinions on Twitter which is now considered a valuable online source for opinions. Therefore, sentiment analysis on Twitter is a rapid and effective way of gauging public opinion about different topics. Thus, in this work, we report on users' perceptions and experiences of using different teleconference applications (i.e., Zoom, Skype, Microsoft Teams, Google Hangouts, HouseParty, Google Duo, Jitsi.). Hence, as detailed in Sec.~\ref{sec:scraping}, we scraped tweets from Twitter using different keywords to analyze the different problems users face such as quality of audio and video calls, reliability, the maximum number of people the application supports (i.e., one to one or group video chats), hijacking video conversations by uninvited parties to disrupt the usual proceedings, passing data to third parties, and different technical difficulties. Then, we classify the extracted tweets into positive and negative. Finally, the most important features are extracted and statistical methods are applied to analyze the challenges, strengths, bottlenecks and best ways of using different teleconferencing applications. 

\section{Tweet-Scraping and Sentiment Analysis} \label{sec:scraping}

Recently, more and more people are connecting with social networks. One of the most widely used social networks is Twitter \cite{twitter_stat}. The huge amount of information found in Twitter is attracting the interest of many people since this huge source of information can be used to analyze the public opinion on different topics \cite{alec2009twitter}. Sentiment analysis, also known as opinion mining, has developed many algorithms to identify whether an online text is subjective or objective, and whether any opinion expressed is positive or negative \cite{pang2008opinion}. Sentiment analysis has been successfully used for prediction and measurement in a variety of domains, such as stock market, politics and social movements using Twitter data \cite{bollen2011Twitter,choy2011sentiment,tumasjan2010predicting}. Thus, we extracted tweets about different teleconference and remote work applications using keywords relevant to teleconferencing applications as shown in Table \ref{table:sentiment_percentage}, and classified these tweets into positive and negative. Note that these are the maximum number of tweets we could extract for each keyword. 

\begin{figure}[ht]
\centering
\includegraphics[width=0.8\columnwidth]{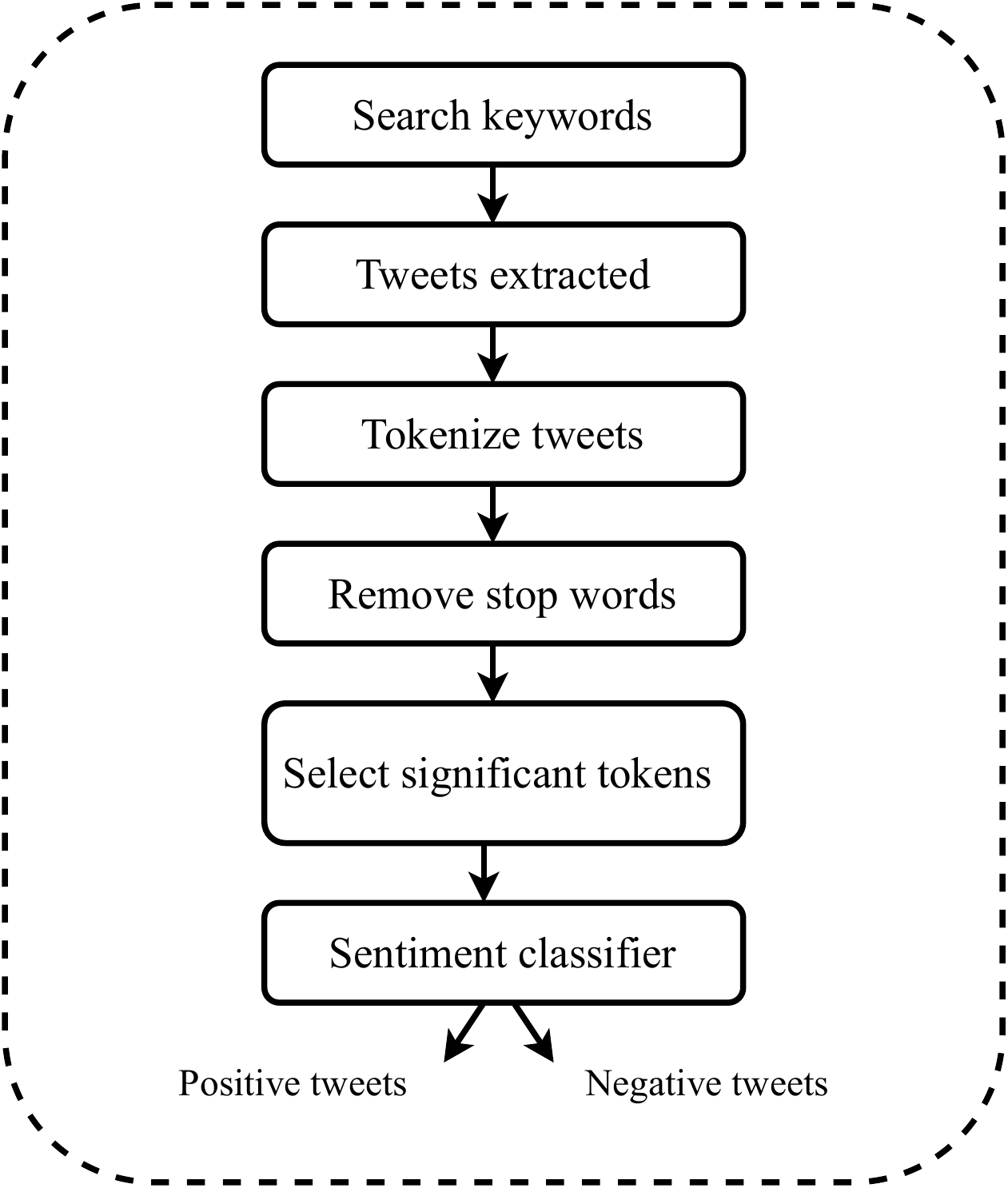}
\caption{Workflow in sentiment analysis.}
\label{fig:graphical_abstract}
\end{figure}

After we extracted the tweets using each of the keywords mentioned in Table~\ref{table:sentiment_percentage}, we analyze and search for the most important features of each teleconference application. Since the extracted raw tweets contain unformatted text, it is important to remove these unimportant texts and characters (i.e., hyperlinks, special characters, emojis, etc.), tokenize the tweets and remove stop words \cite{saif2014stopwords,duncan2015neural}. Afterwards, we select only the significant tokens like adjectives, adverbs, etc. Finally, we passed the tokens to a sentiment classifier to classify them as positive and negative.

Once the tweets are in proper format for analysis, we have performed various types of statistical analysis such as the percentage of positive and negative tweets for each keyword and the most important features of teleconference applications. After the classification of the tweets into positive and negative, we again clean up the tweet text including differences in case (e.g. upper, lower) that will affect unique word counts and remove words that are not useful for the analysis. Finally, we summarize and count the different words found in tweets to extract the most important features of tweets. The selected features are used to analyze the strength and weakness of different teleconference applications.

\section{Methods}

Twitter has a total number of 330 million daily active users \cite{twitter_stat}. People use Twitter to share their opinions and share information for different purposes. The availability of huge data on Twitter has enabled researchers to use its data for different applications. Therefore, we have also used Twitter data to analyze users' perceptions and experiences of using different teleconference applications. 

The sentiment analysis for the extracted tweets is illustrated in Figure \ref{fig:graphical_abstract}. We have totally extracted 410573 tweets using the Python library Tweepy \cite{roesslein2009tweepy}. The retweets and replies were filtered out while collecting the tweets to avoid duplication of the tweets. The data comprised 410,573 English‐language tweets posted between March 17, 2020 and March 29, 2020. Once the tweets are extracted, we calculate for each word its frequency. The words were then listed in decreasing order to find the list of words with the biggest spikes of interest. 

The Cross-Lingual Sentiment (CLS) dataset \cite{Prettenhofer2019} comprises about 800,000 Amazon product reviews for four languages: English, German, French, and Japanese. We used $44584$ English product reviews from this dataset to train the sentiment analysis system. Thus, the extracted tweets using different keywords are classified as positive or negative using the classifier model trained with this CLS dataset.

\section{Analysis}

The analysis of the extracted tweets is discussed in three categories. Firstly, we compare the percentage of positive and negative sentiments for each tweet keyword and teleconference application. Secondly, we discuss the important features extracted from the tweet keywords and teleconference applications. Finally, we compare the percentages of positive and negative tweets of the most important features.

Table \ref{table:sentiment_percentage} depicts the percentage of positive and negative sentiments for each tweet keyword and teleconference application. From the table, we can see that the percentage of positive sentiments outweighs the percentage of negative sentiments for each tweet keyword and teleconference applications. The percentage of positive tweets is the highest for webinar (i.e., 95.51\%) and the lowest for FaceTime (i.e., 59.41\%). The table also shows that Microsoft Teams received the most positive sentiment among videoconferencing applications studied. The difference in the percentage of positive sentiments for remotework (i.e., 90.45\%) and remote work (i.e., 78.49\%) could be remotework is considered a hashtag and the tweets might not contain a direct opinion on the topic (i.e., the tweet is just about a related topic that is less likely to be negative). Generally, from Table \ref{table:sentiment_percentage}, we can conclude that despite the problems of teleconference applications such as audio/video quality, reliability, software/hardware issues, privacy, security and the like, people still have positive opinions for teleconference applications (i.e., average positive sentiment of 71.67\%). Note that since the advertisements in the tweets are classified as positive, there is a small bias in the percentage of positive tweets.

\begin{table}[h]
\caption{The total number of tweets and the percentage of positive sentiments for different keywords.}
\centering
\begin{tabular}{|c|c|c|}
\hline
 \cellcolor[HTML]{9B9B9B}\textbf{Keyword}                &
 \cellcolor[HTML]{9B9B9B}\textbf{Number of tweets}                &
 \cellcolor[HTML]{9B9B9B}\textbf{\% of positive} \\ \hline
Teleconference                      &       10,000  & 70.18                                      \\ \hline
Remotework                          &       11,125  & 90.45                                 \\ \hline
Remote work                         &       23,667  & 78.49                                 \\ \hline
Video conference                    &       25,987  & 71.92                                       \\ \hline
Webinar                             &       100,000 & 95.51                                     \\ \hline
Jitsi                               &       1238   & 63.49                                          \\ \hline
Google Duo                          &       1479 & 70.72                                             \\ \hline
Google Hangouts                     &       10,079 & 62.22                                         \\ \hline
HouseParty                          &       12,160 & 66.75                                       \\ \hline
Microsoft Teams                     &       14,462 & 72.92                                       \\ \hline
Skype                               &       34,849 & 62.88                                         \\ \hline
FaceTime                            &       67,293 & 59.41                                        \\ \hline
Zoom                                &       98,234 & 66.80                                        \\ \hline

\textbf{Total}                & \multicolumn{2}{l|}{\, \, \,  \, \,  \,  \textbf{410, 573}}  \\ \hline
\textbf{Average}                & \multicolumn{2}{c|}{\, \, \, \,  \,  \,  \,   \,  \,  \,  \,  \, \,  \,  \,  \,  \,  \,  \,  \,  \,  \, \textbf{71.67}}  \\ \hline
\end{tabular}
\label{table:sentiment_percentage}
\end{table}

\begin{figure}[h]
\centering
\includegraphics[width=\columnwidth]{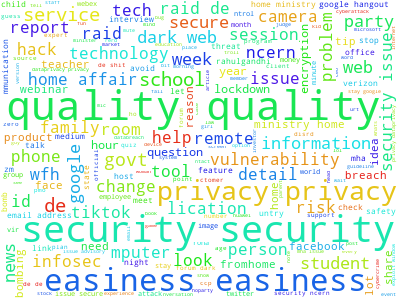}
\caption{The word cloud for the tweet keyword Zoom. The word cloud is generated using 10000 tweets.}
\label{fig:word_cloud}
\end{figure}

To get insight into the reasons behind people's positive and negative sentiments, we further want to identify the characterizing features of each keyword and teleconferencing application. The most important features are extracted by listing the words with their respective frequency after removing the stop words and lemmatization. The grouping of important features in Table~\ref{table:important_features} is selected as follows: the words privacy and private in the tweets will be grouped into the ``privacy" category. Similarly, we use the words secure and security to associate them into the ``security" category. To group tweets into the ``quality" category, we search the words quality, bad and good. Finally, we use the words easy and hard to aggregate them into the ``easiness" category. Easiness includes seamless use of applications such as attendees can join by a publicly shared link from anywhere, and joining does not require downloading any software. Figure~\ref{fig:word_cloud} displays the word cloud (i.e., a visual representation showing the most relevant words) of 10000 tweets for Zoom keyword. The Figure shows that quality, easiness, security and privacy are the most frequently used keywords in descending order.

Table~\ref{table:important_features} reveals that privacy, security, quality and easiness are the most tweeted features for each tweet keyword and teleconference application. Zoom has the highest number of privacy and security tweets from all teleconference applications. This may be due to the recent reports of security and privacy problems of Zoom \cite{zoom_security}. People however seem to be content with the audio and video quality features of Zoom. 

\begin{table*}[h!]
\caption{Occurrences of tweets in four largest categories for each keyword.}
\centering
\begin{tabular}{|c|c|c|c|c|c|}
\hline
 \cellcolor[HTML]{9B9B9B}\textbf{Keyword}                &
 \cellcolor[HTML]{9B9B9B}\textbf{Privacy} & 
 \cellcolor[HTML]{9B9B9B}\textbf{Security} &
 \cellcolor[HTML]{9B9B9B}\textbf{Quality} &
 \cellcolor[HTML]{9B9B9B}\textbf{Easiness} &
  \cellcolor[HTML]{9B9B9B}\textbf{Total} \\ \hline
Teleconference & 234 & 600 & 308  & 143  & 1285  \\ \hline
Remotework & 60 & 927 & 392 & 260 & 1639 \\ \hline
Remote work & 154 & 1400 &1034 & 809 & 3397 \\ \hline
Videoconference & 439 & 1036 &1191 & 383 & 3049 \\ \hline 
Webinar & 635 & 2575 &3404 & 1138 & 7752 \\ \hline
Jitsi  & 73 & 129 &107 & 35 & 344 \\ \hline
Google Duo & 6 & 24 &295 & 18 & 343 \\ \hline
Google Hangouts  & 78 & 109 &686 & 166 & 1039 \\ \hline
HouseParty & 72 & 93 &378 & 101 & 644 \\ \hline
Microsoft Team & 127 & 996 & 431 & 215 & 1769 \\\hline
Skype & 564 & 87 & 506 & 144 & 1301 \\\hline
FaceTime & 184 & 113 & 3001 & 935 & 4233 \\\hline
Zoom & 708 & 1387 &3162 & 1613 & 6870 \\\hline
\textbf{Total} & 3100 & 8876 &14587 & 5817 & 32380 \\\hline
\end{tabular}
\label{table:important_features}
\end{table*}

After the extraction of the four most important features in the tweets, we again classify the opinions of users of these features into positive and negative. Thus, as shown in Table~\ref{table:sentiment_features}, Microsoft Team receives the highest positive sentiment in category ``easiness" (80.0\%). The contributing factors could be its full integration with Microsoft 365 that enables calls to be easily scheduled. In addition to this, users can join from their web browser without downloading Microsoft Teams. Zoom looks to be the choice for quality feature for many users. The contributing factors could be that it works reasonably well with few dropouts and high-quality video that seamlessly shifts to show whoever is speaking. The low sentiment percentage of Zoom about its privacy feature (i.e., 55.6\%) is related to its privacy issues in recent months. Hence, from the results of Table~\ref{table:sentiment_features}, we can see that although users are skeptical about the privacy, security, quality and easiness features of different teleconference applications, they still have positive opinions about them. The average sentiment of positive tweets is greater than negative sentiments for the four features. 

The occurrence of privacy and security words in the tweets could be because of the recent negative news about privacy and security problems of some teleconference applications. Easiness of the teleconference application is also another important feature for many users since some teleconference applications do not require downloading any software to join the video/audio conference call. Video and audio qualities are also a concern for many users since the quality of teleconference applications has reduced because of the high demand of teleconference applications throughout the world. 

\begin{table}[ht!]
\caption{The percentage of positive tweets for the four most occurring target keywords.}
\centering
\begin{tabular}{|p{2.23cm} | p{0.85cm} | p{0.945cm} | p{0.85cm} |p{0.955cm} | }
\hline
 \cellcolor[HTML]{9B9B9B}\textbf{Keyword} &
 \cellcolor[HTML]{9B9B9B}\textbf{Privacy} & 
 \cellcolor[HTML]{9B9B9B}\textbf{Security} &
 \cellcolor[HTML]{9B9B9B}\textbf{Quality} &
 \cellcolor[HTML]{9B9B9B}\textbf{Easiness} \\ \hline
Teleconference & 53.8\% & 66.8\%  & 72.1\%  & 31.5\%  \\ \hline
Remotework & 88.3\% & 90.3\% & 90.3\% & 88.1\% \\ \hline
Remote work & 78.6\% &90.9\% & 77.0\% & 78.4\% \\ \hline
Videoconference & 69.2\% &75.5\% & 72.0\% & 69.7\% \\ \hline 
Webinar & 95.7\% &97.0\% & 95.5\% & 94.6\% \\ \hline
Jitsi & 78.9\% &69.7\% & 56.1\% & 68.6\% \\ \hline
Google Duo & 50.0\%&54.1\% & 86.7\% & 72.2\% \\ \hline
Google Hangouts & 59.0\% &58.7\% & 52.9\% & 60.2\% \\ \hline
HouseParty & 50.0\% &36.6\% & 74.3\% & 73.3\% \\ \hline
Microsoft Team & 57.5\% & 59.8\% & 68.2\% & 80.0\% \\\hline
Skype & 95\% & 48.2\% & 61.2\% & 53.4\% \\\hline
FaceTime & 57.1\% & 49.6\% & 61.9\% & 36.8\% \\\hline
Zoom & 55.6\% &60.6\% & 83.7\% & 28.7\% \\\hline
\textbf{Average} & \textbf{68.36\%}  & \textbf{65.98\%} & \textbf{73.22\%} & \textbf{64.27\%} \\\hline
\end{tabular}
\label{table:sentiment_features}
\end{table}

To extract deeper understanding of people's experience with teleconferencing applications, we further posted questions on Twitter to get direct responses from users. A central difference is that where tweet scraping gives spontaneous and unsolicited statements about teleconference use, by directly asking for opinions, people think specifically about teleconferencing. Our expectation was that this way we would get answers of different characteristics. Unfortunately, our questions did not receive enough responses that their analysis would be worthwhile. 

\section{Discussion}

As Corona-virus locked-downs have moved many in-person activities online, the use of the video-conferencing platform has quickly escalated. So, too, have concerns about security, reliability, media quality, easiness, reliability, capacity and technical difficulties. In addition to these, some teleconference application company’s privacy policies allow them to collect all sorts of personal data. 

Thus, we would advise users of teleconference applications to apply the following practices to minimize the above mentioned problems: password-protect meetings, consider advanced features such as using waiting rooms and locking the meeting room once all attendees have arrived, do not share meeting ID, and consider using Single Sign-On (SSO) technologies such as Google/Okta for login. Users also need to make sure that the video conferencing application in use is patched with the latest vendor-provided updates and have automated upgrades turned on. In addition to these, users should  review and enable appropriate security and privacy settings to prevent malicious actors from exploiting known vulnerabilities. Finally, lack of sufficient bandwidth and other network-related issues are the one of the causes for audio and video quality problems. Thus, users need to check their Internet connection speed and the bandwidth requirements of the video conferencing application in use. The common video conferencing problems and their solutions are also outlined in \cite{video_conference_problems_solutions_1,video_conference_problems_solutions_2}.

\section{Conclusions}

The demand of teleconference applications has seen a huge rise in downloads since quarantines were imposed in March 2020 around the world. They are now  used by millions of users around the world for work, education, personal and social purposes. However, the high demand of teleconference applications created some dissatisfaction among users such as privacy, security, quality of service, reliability and different software and hardware technical issues. Thus, in this work, we have analyzed users' perception towards different teleconference applications using Twitter posts. We apply sentiment analysis on the extracted tweets to classify them as positive and negative, and extract the most important features both from the positive and negative tweets. Based on these features, we highlighted the strengths, weaknesses and challenges of different teleconference applications, and propose the best ways of using them.

Overall, however, our analysis has already shown that users post mostly positive comments about teleconferencing applications on Twitter. This suggests that users are mostly content with the performance of teleconferencing applications. The most prominent development objectives, however, for teleconferencing applications are, according to our analysis, quality, security, easiness and privacy.

This research was conducted during a relatively small window in time. Thus, a total of only 410,573 tweets were extracted using selected keywords and teleconference applications. In addition to this, the tweets were extracted only for English language. Thus, the future research could focus on extracting many tweets for different languages, and classify the strengths, weaknesses and challenges of different teleconference applications for each language.  The future research could also take geographical regions into account when tweets are scraped and analyzed. In addition to this, the future work could include collecting direct responses using questionnaires from users of teleconference applications to get filtered opinions.




\end{document}